      [1999/12/01 v1.4c Il Nuovo Cimento]
\documentclass{cimento}
\usepackage{graphicx}
\usepackage{amssymb}
\title{Extreme gravitational lensing by supermassive black holes}
\author{V.~Bozza\from{ins:x}}
\instlist{\inst{ins:x} Dipartimento di Fisica ``E.R. Caianiello'',
Universit\`a di Salerno, via S. Allende, Baronissi (SA), Italy.
Istituto Nazionale di Fisica Nucleare, Sezione di Napoli, Italy.}
\PACSes{\PACSit{95.30.Sf,04.70.Bw,98.62.Sb}{}}
\begin{document}

\maketitle

\begin{abstract}
Extreme gravitational lensing refers to the bending of photon
trajectories that pass very close to supermassive black holes and
that cannot be described in the conventional weak deflection
limit. A complete analytical description of the whole expected
phenomenology has been achieved in the recent years using the
strong deflection limit. These progresses and possible directions
for new investigations are reviewed in this paper at a basic
level. We also discuss the requirements for future facilities
aimed at detecting higher order gravitational lensing images
generated by the supermassive black hole in the Galactic center.
\end{abstract}

\section{Introduction}

All known astrophysical cases of bending of photon trajectories by
gravitational fields can be interpreted using the conventional
weak deflection paradigm, originally formulated by Einstein
\cite{Ein}. Yet it is clear that photons passing very close to
black holes suffer very large deflections, which must be addressed
in a full general relativistic context. However, integrating null
geodesics in full general relativity usually leads to very
complicated analytical formulae or heavy numerical codes, which
typically obscure the basic physical interpretation. The Strong
Deflection Limit (SDL) is an analytical tool that allows to derive
simple analytical formulae describing the higher order images
appearing in extreme gravitational lensing. In this work we review
the main progresses achieved in the recent years in this technique
and its application to the most interesting physical case: the
black hole in the Galactic center (Sgr A*). All relevant formulae
derived in previous works are recalled and restated here in the
simplest possible form.

This work is structured as follows: \S~2 explains the basic
phenomenology of extreme lensing; \S~3 introduces the SDL for
spherically symmetric black holes; \S~4 generalizes the method to
spinning black holes; \S~5 applies the formulae to Sgr A*,
examining possible sources for extreme gravitational lensing; \S~6
generalizes the method to sources very close to the black hole;
\S~7 contains the conclusions.

\section{Basic phenomenology}

Performing an exact integration of the null geodesics equation in
Schwarzschild metric, Darwin found a general expression for the
deflection angle in a Schwarzschild geometry without any
approximation \cite{Dar}. This formula looks quite complicated,
since it involves elliptic integrals. Nevertheless, taking the
Weak Deflection Limit (WDL) for photon passing far enough from the
black hole, it reduces to the classical Einstein's formula for the
deflection angle
\begin{equation}
\alpha(u)= \frac{4GM}{c^2 u},
\end{equation}
where $u$ is the impact parameter of the photon as emitted by the
source, $M$ is the mass of the black hole, $c$ is the speed of
light and $G$ is the Newton constant.

\begin{figure}
\includegraphics{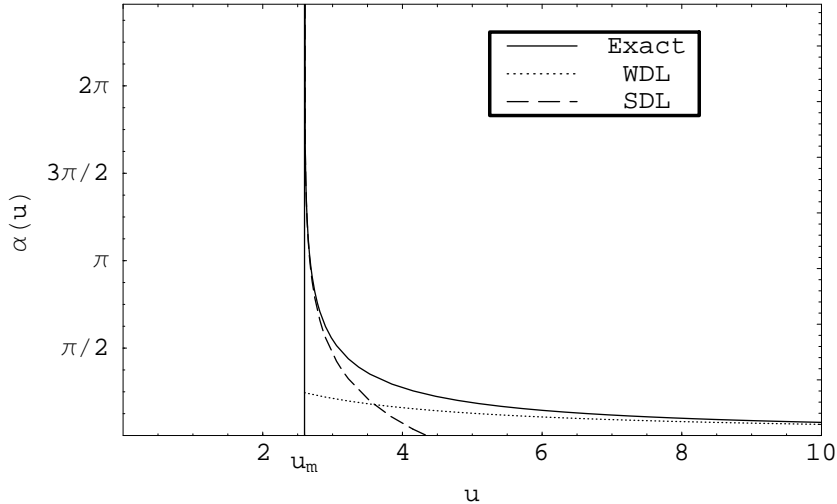}
\caption{Comparison between the exact deflection angle for photons
in a Schwarzschild metric and the two limits discussed in the
text. The impact parameter is in units of the Schwarzschild
radius.} \label{Fig alpha}
\end{figure}

Fig. \ref{Fig alpha} contains a comparison between the exact
deflection angle and the WDL approximation, along with the SDL
approximation to be introduced in the following section. The exact
deflection angle and the WDL perfectly agree at impact parameters
much larger than the Schwarzschild radius of the black hole.
However, as the impact parameter becomes comparable to the
Schwarzschild radius, the approximations underlying the derivation
of the WDL formula are no longer valid. As we lower the impact
parameter $u$, the deflection grows larger and larger, until it
diverges at a limiting value of the impact parameter, indicated by
$u_m$. For the Schwarzschild black hole, we have $u_m=3\sqrt{3}
GM/c^2$.

Very large deflections are possible for $u$ close to $u_m$. A
deflection of order $\pi$ means that the photon turns around the
black hole and goes back to the source. The possibility of such a
back-scattering was already examined by Luminet \cite{Lum} and was
recently revived with the name of retro-lensing \cite{HolWhe}.

A deflection of order $2\pi$ is experienced by photons that
perform a complete loop around the black hole before emerging in
the direction opposite to the source. Higher deflections are
possible for values of $u$ closer and closer to $u_m$,
corresponding to photons performing any number of loops around the
black hole before emerging in any direction.

Photons with impact parameter exactly equal to the minimum value
$u_m$ would be injected on a circular orbit with fixed radius
$r_m$ around the black hole. The spherical surface of radius
$r_m$, containing these circular photon orbits, is also called
photon sphere. In Schwarzschild metric, the radius of the photon
sphere is $r_m=3GM/c^2$. Photons with impact parameter $u<u_m$
just fall inside the event horizon and are definitely lost.

The divergence of the deflection angle for $u \rightarrow u_m$ is
a genuine feature of the strong gravitational fields generated by
black holes. Since it is predicted by general relativity, one can
wonder whether possible observable effects of this divergence
exist.

Let us define the optical axis as the line joining the observer
with the black hole. If we have a source lying behind the black
hole, classical gravitational lensing theory states that the
observer sees two images formed by photons weakly deviated by the
black hole. These two images are formed by photons passing far
from the black hole, which can be usually described in the WDL.
Besides these two main images, the fact that there are values of
$u$ such that $\alpha=2\pi$ implies that there exist some photons
passing very close to the black hole that perform one complete
loop around the black hole and finally reach the observer. These
photons form an additional pair of images (one on each side of the
black hole). Furthermore, there are photons performing two loops,
giving rise to one more additional pair, and so on. The divergence
in the deflection angle as $u \rightarrow u_m$ is thus responsible
for the existence of two infinite sequences of higher order images
appearing very close to the black hole. The angular separation
from the black hole of images formed by photons with impact
parameter $u$ is simply given by $\theta=\arcsin
(u/D_\mathrm{OL})\simeq u/D_\mathrm{OL}$, with $D_\mathrm{OL}$
being the distance between observer and lens.

No image of distant sources can appear at angular separations less
than $\theta_m= u_m/D_\mathrm{OL}$, because no deflection is
possible with $u<u_m$. The minimum angle $\theta_m$ defines the
angular radius of the so-called shadow of the black hole
\cite{Dar,Lum}. The higher order images would appear just slightly
outside the border of this shadow.

A perfect alignment of source, lens and observer on the same line
would cause each pair of images to merge together and form an
Einstein ring. Besides the classical WDL Einstein ring, we would
have an infinite sequence of higher order Einstein rings just
outside the shadow of the black hole. With respect to standard WDL
gravitational lensing, extreme black hole lensing also opens the
possibility of forming retro-lensing Einstein rings, i.e. ring
images generated by a source exactly in front of the black hole
whose photons turn around the black hole and then reach the
observer.

The basic phenomenology of higher order images generated by a
Schwarzschild black hole was described by Darwin in 1959
\cite{Dar}. However, higher order images were practically
forgotten until Virbhadra and Ellis noticed that higher order
images could become observable around the supermassive black hole
in the Galactic center and perhaps around central black holes of
nearby galaxies \cite{VirEll}. After their work, higher order
images have been the object of specific studies by numerous
authors both numerically and analytically.

\section{Gravitational lensing in the Strong Deflection Limit:
spherically symmetric black holes} \label{Sec Spheric}

The study of higher order images by numerical methods is quite
complicated because it is necessary to integrate null geodesics
with very high accuracy. On the other hand, Darwin himself gave an
expansion of his exact deflection formula that is valid very close
to the divergence in $u_m$. It reads
\begin{equation}
\alpha(u)=-c_1 \log \left(\frac{u}{u_m}-1 \right) +c_2 + O\left(u
- u_m \right), \label{SDL}
\end{equation}
with the coefficients $c_1=1$ and $c_2=\log
\left[216(7-4\sqrt{3})\right]-\pi$ in Schwarzschild metric.

This formula approximates the divergence in $u_m$ very well, as
can be seen in Fig. \ref{Fig alpha}. Therefore, it provides an
excellent and highly simplified starting point for the analytical
treatment of higher order images. In particular, the angular
position and the magnification of the higher order images assume
extremely simple expressions as functions of the position of the
source \cite{Dar,Oha}. They read
\begin{eqnarray}
&& \theta_n=\theta_m\left(1+e^{(c_2-\phi_s-2n\pi)/c_1} \right) \label{thetan}\\
&& \mu_n=
\left(\frac{D_\mathrm{OS}}{D_\mathrm{LS}}\right)^2\frac{\theta_m^2}{c_1
\sin \phi_s}e^{(c_2-\phi_s-2n\pi)/c_1}, \label{mun}
\end{eqnarray}
where $D_\mathrm{LS}$ is the distance from the lens to the source,
$D_\mathrm{OS}$ is the distance from the observer to the source,
$\phi_s$ is the azimuth of the source measured in a reference
frame centered on the black hole ($\phi_s=0$ means source aligned
behind the black hole), and $n=1,$ $2,$ $3,$ $\ldots$ identifies
the order of the image.

Since the higher order images are generated in a strong gravity
regime, they have the potentiality to become a test of general
relativity outside the weak field approximation. In principle they
can be used to distinguish general relativity from alternative
theories of gravity that lead to the same expectations in weak
field tests but different predictions in a strong gravity regime.
However, the first step is to understand whether they are just an
exceptional outcome of general relativity or they are generic to
all theories of gravity.

Ref. \cite{Boz1} demonstrates that the deflection angle formula
(\ref{SDL}) is universal, in the sense that it is exactly the same
for any spherically symmetric metric admitting a static limit
(i.e. a spherical surface where $g_{tt}=0$, with $t$ being the
proper time for an observer at infinity). So, whatever the theory
of gravity we consider, the black hole metric will always be
characterized by a photon sphere and thus by a logarithmic
divergence in the deflection angle. What changes from one metric
to another is the numerical value of the coefficients $u_m$,
$c_1$, $c_2$. These three coefficients are a sort of identity card
of the black hole metric which is inherited by the higher order
images. In principle, since the position and the luminosity of the
higher order images depend on the three coefficients in the SDL
formula, by the observation of the higher order images it is
possible to identify the correct black hole metric and thus the
correct theory of gravity.

This possibility opened by higher order images has pushed many
authors \cite{Spheric} to calculate the SDL coefficients for
several black hole metrics using the general method outlined in
Ref. \cite{Boz1}. This method can be easily extended to the
retro-lensing geometry \cite{EirTor,BozMan}, taking $\phi_s=\pi$
and $n$ starting from 0 in Eqs. (\ref{thetan}) and (\ref{mun}).
Time delay calculations have been done in Ref. \cite{TimDel}.
Higher orders in the expansion (\ref{SDL}) were calculated in Ref.
\cite{IyePet} for the Schwarzschild black hole.

It is important to note that in the intermediate region $[0.15<
\alpha <2]$ both the WDL and the SDL approximations reproduce the
exact deflection angle with an accuracy worse than 10$\%$. The
only known analytical approximations able to cover this region are
based on variational methods \cite{Amore}.

\section{Gravitational lensing in the Strong Deflection Limit:
spinning black holes} \label{Sec Kerr}

The integration of null geodesics in the Kerr black hole metric is
considerably simplified by the separability of the Hamilton-Jacobi
equation \cite{Car}. The geodesics equations can be put in the
form of first integral equations, which can be solved in terms of
elliptic functions. The explicit form of these analytical
solutions is extremely complicated, but can be exploited in
building fast and efficient numerical codes aimed at specific
aspects of the propagation of photons in the Kerr metric (see e.g.
\cite{CunBar,RauBla}). In spite of the large number of studies in
this field, the phenomenology of higher order images was still
poorly understood before the techniques of the SDL came into play.

There are some fundamental differences between extreme lensing by
spinning black holes and spherically symmetric black holes.

Eq. (\ref{mun}) shows that when the source is perfectly aligned on
the optical axis ($\phi_s=0$) the magnification diverges for all
higher order images (the same happens for the two classical WDL
images). Within gravitational lensing theory, this divergence
signals the fact that there is a unique caustic point lying on the
optical axis for all images. When a source approaches this point,
all images become simultaneously bright and finally degenerate
into Einstein rings.

In spinning black holes, there is one separate caustic for each
pair of images. All caustics are shifted from the optical axis in
the sense opposite to the rotation of the black hole. The shift
becomes larger and larger with the order of the images. As a
consequence, if the source is close to the WDL caustic, only the
WDL images will become bright while all the others stay faint. If
the source is close to the first order SDL caustic, only the first
order images get bright while all the others stay faint, and so
on.

In addition to the shift from the optical axis, the caustics are
no longer pointlike, but assume the classical shape of the
4-cusped astroid, which is typical of the breaking of spherical
symmetry in gravitational lensing. When the source is inside the
caustic, an additional pair of images appears.

In a series of papers devoted to spinning black holes
\cite{BozEq,VazEst,KerObs,KerGen} the SDL paradigm has been
gradually extended to gravitational lensing by spinning black
holes. Here we report the most recent results in Ref.
\cite{KerGen}.

In order to keep all the results analytical, we can expand all
formulae to second order in the black hole spin $a$, finding very
accurate results up to $a=0.1$ (in our notations, $a=0.5$ is the
extremal Kerr black hole). In this way, expressions for the
position and the shape of the caustics can be explicitly derived.
In a reference frame centered on the black hole with polar axis
defined by the projection of the spin axis on the observer sky,
the azimuthal shift from the optical axis is
\begin{equation}
\Delta_k=  -4\left[\frac{k\pi}{3\sqrt{3}} + \log(2
\sqrt{3}-3)\right]a \sin i. \label{shift}
\end{equation}
where $i$ is the inclination of the spin axis with respect to the
observer line of sight and $k$ is the caustic order, being $3$,
$5$, $7$, $\ldots$ for higher order images in standard lensing
configuration and $2$, $4$, $6$, $\ldots$ for retro-lensing
configuration. Note that $\Delta_k$ is always negative, which
means that the shift is in the clockwise sense.

The shift in the altitude angle is much smaller, being of second
order in $a$:
\begin{equation}
\delta_k  = (-1)^k\frac{1}{2} \Delta_k^2 \cot i
\end{equation}

The semi-amplitude of the astroid caustic is
\begin{equation}
R_k=\frac{2}{9} (5k\pi+8\sqrt{3}-36)a^2 \sin^2 i. \label{deltacau}
\end{equation}

\begin{figure}
 \centering
 \resizebox{10cm}{!}{\includegraphics{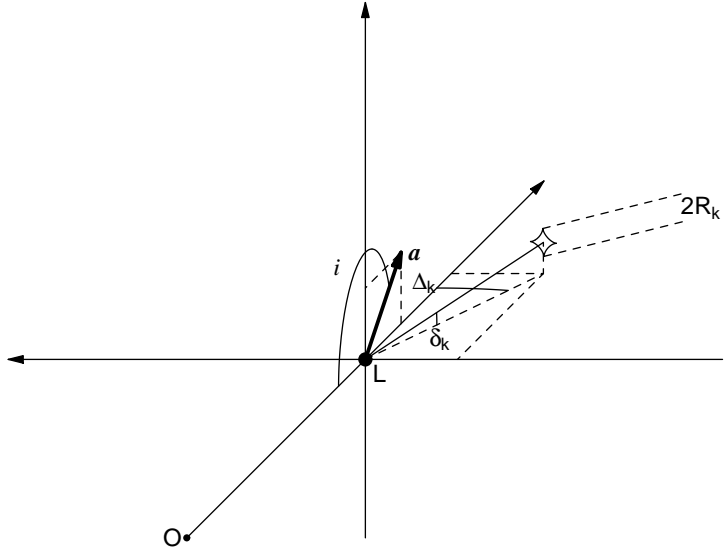}}
 \caption{Position and shape of the caustics in Kerr gravitational lensing.
 The observer is in O, the black hole is in L, with spin oriented along the vector \textbf{\textit{a}}.
 The inclination of the spin with respect to the line of sight is $i$. The azimuth of the
 center of the caustic in this reference frame is given by $\Delta_k$ and its altitude is $\delta_k$.
 The full amplitude of the astroid caustic is $2R_k$.} \label{Fig cau}
\end{figure}

From these equations it is evident that the shift and the
extension of the caustic increase with the black hole spin, its
inclination and the caustic order. The precise geometric meaning
of these quantities is illustrated in Fig. \ref{Fig cau}.

The position of the images for a source close to a caustic (which
is the most relevant case) can be found in two steps. First, the
equation
\begin{equation}
\delta \vartheta_s \cos \psi + (-1)^k (\delta \phi_s -R_k \cos
\psi)\sin \psi = 0, \label{Eqimages}
\end{equation}
must be solved for the angle $\psi$. Here $\delta \phi_s$ and
$\delta \vartheta_s$ represent the displacement of the source from
the center of the caustic in the azimuth and polar angle
respectively. With reference to Fig. \ref{Fig cau}, if $\delta
\phi_s>0$ the source is on the left of the caustic; if $\delta
\vartheta_s>0$ the source is below the caustic.

For each real solution of Eq. (\ref{Eqimages}) we have a distinct
image. The coordinates of the images in the observer sky are then
simply given by
\begin{eqnarray}
&&\theta_{1,k}= \theta_m\left( 1+ \epsilon_k \right) \cos \psi \\
&&\theta_{2,k}= \theta_m\left( 1+ \epsilon_k \right) \sin \psi,
\end{eqnarray}
with $\epsilon_k=216\left(7-4\sqrt{3}\right)e^{-k\pi}$. With
reference to Fig. \ref{Fig cau}, if $\theta_{1,k}>0$ the observer
sees the image on the right of the black hole; if $\theta_{2,k}>0$
the observer sees the image above the black hole.

The radial and tangential magnifications are respectively given by
\begin{eqnarray}
&& \mu_r= \frac{D_\mathrm{OS}}{D_\mathrm{LS}} \theta_m \epsilon_k \label{mur}\\
 && \mu_t= (-1)^k\frac{D_\mathrm{OS}}{D_\mathrm{LS}} \frac{\theta_m
\cos\psi}{\left(\delta \phi_s- R_k \cos^3 \psi  \right)}.
\label{mut}
\end{eqnarray}

By these very simple formulae, we can describe the whole
phenomenology of higher order images in Kerr black hole lensing,
including caustic crossing, with formation and destruction of
additional pairs of images.

It is interesting to note that, to lowest order in $a$, all
observables in Kerr lensing are functions of $a \sin i$. In
practice, only the projection of the spin on a plane orthogonal to
the line of sight is accessible by gravitational lensing
measurements.

Another interesting consideration regards the stability of higher
order images with respect to external fields. The study of extreme
lensing by a Schwarzschild black hole embedded in an external
uniform gravitational field has shown that the external field
causes the shift and the formation of extended astroid caustics in
a way qualitatively similar to Kerr metric, but with very
different quantitative aspects \cite{BozSer}. In particular, the
caustic extension induced by realistic external fields would be
much smaller than that due to an intrinsic black hole spin.

The SDL has been recently extended to the Kerr-Sen dilaton-axion
spinning black hole \cite{KerSen}.

\section{Sgr A* as an extreme gravitational lens}

The radio source Sgr A* is believed to host a supermassive black
hole with mass $M=3.6\times 10^6$ $M_\odot$ \cite{Eis}. With a
distance from the sun $D_\mathrm{OL}=8$ kpc, the shadow of the
black hole would have an angular radius $\theta_m=23$
microarcseconds. The higher order images would appear just outside
the shadow of the black hole and thus demand a very high angular
resolution to be detected.

Resolutions of this order can be reached with very long baseline
interferometry in the radio cm-band \cite{Kri}. However, at
wavelengths higher than 1 mm, the scattering by electrons
surrounding Sgr A* blurs the image, making any progress in the
resolution useless \cite{Lo}. Sub-mm imaging would probably be
able to show the existence of the shadow of the black hole
\cite{FMA,Zak}.

Higher chances of seeing higher order images exist in the infrared
bands, where the blur is negligible and the dust extinction still
allows good observations of the Galactic Center. In addition to
this, several stellar sources have been identified and accurately
followed through their orbit around the black hole \cite{Eis}.
With these sources, we can make accurate predictions for the
appearance and luminosity of secondary and higher order images
\cite{DeP,BozMan,S1-S14}. Unfortunately, Sgr A* has its own
emission in the IR, which would render the identification of
higher order images definitely challenging. On the other hand,
there are good chances to see a secondary image in an intermediate
regime between the WDL and the SDL.

The best chances to observe higher order images come from the
X-ray band, where several point sources have been detected and
identified as low mass X-ray binaries \cite{Muno}. In this band,
the emission by Sgr A* is rather low, providing a very low noise
to higher order images \cite{KerObs,KerGen}. The chance alignment
of one of these sources behind or in front of the black hole,
would generate observable higher order images. In this
electromagnetic band, resolutions better than 1 microarcsecond
should be achievable by future X-ray interferometers in space
(MAXIM, http://maxim.gsfc.nasa.gov).

\section{Sources close to black holes}

All the formulae in \S~\ref{Sec Spheric} and \S~\ref{Sec Kerr} are
strictly valid for sources very far from the black hole, where
very far means that their distance is much larger than the
Schwarzschild radius. Yet, we know that most black holes are
surrounded by accretion disks and infalling matter, which provide
very interesting sources for gravitational lensing. In order to
include such sources in our formalism, we need to revise the
derivation of the SDL without taking the limit $D_\mathrm{LS} \gg
2GM/c^2$.

Actually, this can be done without too much trouble and the whole
general method of Ref. \cite{Boz1} can be reformulated taking care
of the source and the observer distance \cite{FinDLS}. In
particular, the azimuthal shift becomes\footnote{It is
inappropriate to speak about a deflection angle if the source is
not in the asymptotic region.}
\begin{equation}
\Delta \phi (u)=-c_1\log \left( \frac{u/u_m-1}{ \eta_\mathrm{O}
\,\eta_\mathrm{S}}\right)+c_2+\pi, \label{FinDLS}
\end{equation}
where $\eta_\mathrm{O}=1-r_m/D_\mathrm{OL}$ and
$\eta_\mathrm{S}=1-r_m/D_\mathrm{LS}$. The coefficient $c_1$ is
unchanged, whereas the general expression of $c_2$ is slightly
changed (see Ref. \cite{FinDLS} for details).

The main change from Eq. (\ref{SDL}) to (\ref{FinDLS}) is
represented by the coefficients $\eta_\mathrm{O}$ and
$\eta_\mathrm{S}$ in the argument of the logarithm. Whereas
$\eta_\mathrm{O}$ is significantly different from one only in the
unrealistic situation of an observer very close to the black hole,
$\eta_\mathrm{S}$ can be less than one for sources close to the
black hole and even negative for sources inside the photon sphere.
This is reflected in the formula for the position of the images,
which allows higher order image to form inside the shadow when the
source is inside the photon sphere
\begin{equation}
\theta_n=\theta_m\left(1+\eta_\mathrm{O} \,\eta_\mathrm{S}
e^{(c_2-\phi_s-2n\pi)/c_1} \right).
\end{equation}
The value of $c_2$ in the case of the Schwarzschild black hole is
explicitly
\begin{eqnarray}
&& c_2=-\pi+5\log[6]+b_\mathrm{O}+b_\mathrm{S} \\
&& b_\mathrm{O}=-2\log\left[3+\sqrt{3+\frac{9}{D_\mathrm{OL}}}\right] \\
&&
b_\mathrm{S}=-2\log\left[3+\sqrt{3+\frac{9}{D_\mathrm{LS}}}\right],
\end{eqnarray}
where the distances of observer and source to the black hole are
measured in Schwarzschild radii.

The extension to sources close to the black hole can be also done
for spinning black holes. Here as well everything remains
qualitatively identical to the infinitely distant source case. The
only changes come into the expression of the shift of the caustic,
which is updated to
\begin{equation}
\Delta_k=-\left\{\frac{4k\pi}{3\sqrt{3}}+4\log\left(2\sqrt{3}-3\right)
+\log\left[
\frac{(2\sqrt{D_{LS}}+\sqrt{3+D_{LS}})^2}{9(D_{LS}-1)}\right]
\right\} a \sin i,
\end{equation}
and in the expression of the semi-amplitude of the caustic
\begin{equation}
R_k=\left[\frac{2}{9} (5k\pi+8\sqrt{3}-36)  +\frac{2\left(
9+4D_{LS}-4\sqrt{D_{LS}}\sqrt{3+D_{LS}}\right)}{3\sqrt{3}\sqrt{D_{LS}}\sqrt{3+D_{LS}}}\right]a^2
\sin^2 i.
\end{equation}

The position of the images can be found with the same algorithm
described in \S~\ref{Sec Kerr}. The magnification, instead, is not
well-defined if the source is inside the gravitational field of
the black hole, because there is no reference unlensed image to
compare with. The determination of surface brightness and shape of
the image must take into account the gravitational and Doppler
redshift and is thus better performed on specific source models.

\section{Conclusions}

Extreme gravitational lensing is a very spectacular though elusive
phenomenon, which demands a great effort in order to be observed.
The most promising extreme lens is represented by the supermassive
black hole in the Galactic center, which anyway requires
resolutions of order microarcseconds for the direct observation of
higher order images. Suitable known sources for gravitational
lensing are giant stars in the IR band and low mass X-ray binaries
in the X-ray band.

In this paper we have reviewed recent results on gravitational
lensing in the Strong Deflection Limit, which is an approximation
devoted to the analytical determination of the properties of
higher order images. Within this framework, it has been proved
that the logarithmic divergence in the deflection angle is a
universal feature of all black hole metrics. There exists a
general method to determine the SDL coefficients of the deflection
angle for any given spherically symmetric black hole metric. This
method has been applied to numerous metrics. Higher order images
are formed by photons performing one or more loops around the
black hole before reaching the observer. The position and the flux
ratios of higher order images depend on the specific metric
through the SDL coefficients and are thus able to track any
possible deviations from general relativity in the strong field
regime.

The SDL can be extended to spinning black holes, where several new
features emerge, such as extended and shifted caustics and
additional images. The position and the shape of the caustics can
be expressed by simple analytical formulae and the lens equation
for sources near to caustics, which represent the most physically
relevant situation, can be solved.

The extension of the SDL to the case of sources very close to
black holes opens the way to even more interesting investigations
of extreme gravitational lensing. In fact, whereas direct
observation of higher order images is very difficult and probably
very far to come in the future, the contribution of higher order
images to currently observed phenomena involving sources very
close to black holes is already measurable at present time. Light
curves of flares born in the accretion disk and spectral
measurements of Iron K-lines are heavily influenced by the
presence of higher order images. With the SDL setup in its updated
version, it is possible to study such phenomena, bearing in mind
that the large model dependence remains the main uncertainty in
all theoretical attempts to interpret the complicated physics of
the supermassive black holes environment.

\acknowledgments I wish to thank the organizers of the
Italian-Pakistan Workshop on Relativistic Astrophysics, held at
Lecce University, 20-22 June 2007, for their kind invitation.

\end{document}